\documentclass[11pt,onecolumn,a4paper]{article}
\usepackage{graphicx}
\usepackage{natbib}
\bibpunct{(}{)}{;}{a}{}{,}

\title{Constraining global parameters of accreting black holes\\
       by modeling magnetic flares}

\author{R. W. Goosmann$^{1,2}$, M. Mouchet$^3$, M. Dov{\v c}iak$^1$,\\ 
        V. Karas$^1$, B. Czerny$^4$, G. Ponti$^{5,6}$}

\date{~}

\begin{document}

\maketitle

\vspace{-1.5cm}

\begin{center}
{\scriptsize
\noindent
$^1$~Astronomical Institute, Academy of Sciences, Bo{\v c}n{\' i}~II~1401,
     14131 Prague, Czech Republic\\
$^2$~Observatoire de Paris, Meudon, LUTH, 5 place Jules Janssen,
     92195 Meudon Cedex, France\\
$^3$~Laboratoire ApC, Universit\'e Denis Diderot, 2 place Jussieu,
     75251 Paris Cedex 05, France\\
$^4$~Copernicus Astronomical Center, Bartycka 18, 00-716 Warsaw, Poland\\
$^5$~Dipartimento di Astronomia, Universit\`a di Bologna, Via Ranzani 1,
     40127, Bologna, Italy\\
$^6$~INAF--IASF Bologna, via Gobetti 101, 40129, Bologna, Italy\\
}
\end{center}

\begin{abstract}
\noindent We present modeling results for the reprocessed radiation
expected from magnetic flares above AGN accretion disks. Relativistic
corrections for the orbital motion of the flare and for the curved
space-time in the vicinity of the black hole are taken into
account. We investigate the local emission spectra, as seen in a frame
co-orbiting with the disk, and the observed spectra at infinity. We
investigate long-term flares at different orbital phases and
short-term flares for various global parameters of the accreting black
hole. Particular emphasis is put on the relation between the iron
K$\alpha$ line and the Compton hump as these two features can be
simultaneously observed by the {\it Suzaku} satellite and later by
{\it Simbol-X}.
\end{abstract}

\section{Introduction}

The magnetic flare model is an attractive possibility to explain the X-ray
properties of accreting black holes in Active Galactic Nuclei (AGN). It was
originally suggested by \citet{galeev1979} and developed in several subsequent
papers \citep[e.g.][]{haardt1991,haardt1994}. The flares supposedly result
from reconnection events in the magnetized corona of the accretion disk. They
account for the heating of the corona to a very hot ($10^8$--$10^9$~K) and
optically thin medium. This medium explains the production of the observed
primary X-ray spectrum by Compton up-scattering of soft X-ray photons coming
from the accretion disk.  

The flares may appear with different durations and intensities. The rapid
fluctuations seen in X-ray lightcurves of AGN can be explained by the random
appearance of numerous ``weak'' flares distributed across an accretion
disk. Such a model is described in \citet{czerny2004} and
\citet{goosmann2006a}. It enables to constrain global parameters of AGN by
fitting their {\it rms} variability spectrum. But also ``strong'' (very
intense) flares are occasionally seen in the X-ray lightcurves of some
nearby Seyfert galaxies (MCG-6-30-15, Ponti et al. 2004; NGC~5548,
Kaastra et al. 2004). Such flares can dominate the X-ray spectrum on
time scales of thousands of seconds.   

We present spectral modeling over 2--30~keV of strong X-ray flares
occurring above an AGN accretion disk. We concentrate on the
reprocessed component expected from the illuminated patch underneath
the flare source. Our radiative transfer simulations include
computations of the vertical disk structure and modifications of the
spectrum by general relativistic and Doppler effects.

\section{Model}

We assume a compact flare source elevated to a height $H$ above the
accretion disk and emitting a primary spectrum with $F(E) \propto
E^{-\alpha}$. The primary source produces an irradiated, horizontally
stratified hot spot on the disk; the highest illuminating flux appears
at the spot center and the lowest at the border. We solve the
radiative transfer for five different concentric annuli of the hot
spot using the codes {\sc Titan} and {\sc Noar}
\citep{dumont2000,dumont2003}. The vertical structure of the accretion
disk is computed with a new version of the code by
\citet{rozanska2002}. Details of these calculations can be found in
\citet{goosmann2006b}. The resulting local spectra are obtained at
twenty different emission angles. These spectra are then used with the
ray-tracing code {\sc KY} \citep{dovciak2004a,dovciak2004b} that
computes the spectral evolution seen by a distant observer. The code
takes into account all effects of general relativity and the Keplerian
motion of the hot spot around the black hole.

\section{Spectral appearance at different orbital phases}

First, we investigate the observed spectrum of the spot at different
orbital phases. The following parameters for the accreting black hole
are set: mass $M = 10^8 \, M_\odot$, dimensionless spin $a/M = 0.998$,
and accretion rate (in units of the Eddington accretion rate) $\dot m
= 0.001$; for the flare we set $\alpha = 0.9$, $H = 0.5 \, R_{\rm g}$,
with $R_{\rm g} = \frac{GM}{c^2}$, and the spot radius $R_{\rm X} = H
\tan{\theta_0}$, where $\theta_0 = 60^\circ$ is the half-opening angle
of the ``beaming cone''. The thermal flux from the disk at the spot
center, $F_{\rm disk}$, is assumed to be much weaker than the flare
irradiation $F_{\rm X} = 144 \times F_{\rm disk}$. The local radiative
transfer calculations are conducted for a spot at a disk radius of $R
= 7 \, R_{\rm g}$. When computing the distant spectra we neglect that
the local spectra change for other disk radii -- we thus only emphasize
the relativistic effects. We assume that the hot spot exists for a
quarter of an orbit and we analyze the resulting time-integrated
spectrum at four different azimuthal phases. The viewing direction of
the distant observer is inclined by $i = 30^\circ$ with respect to the
disk normal.

\begin{figure}[h!t]
  \centering
  \includegraphics[width=\textwidth]{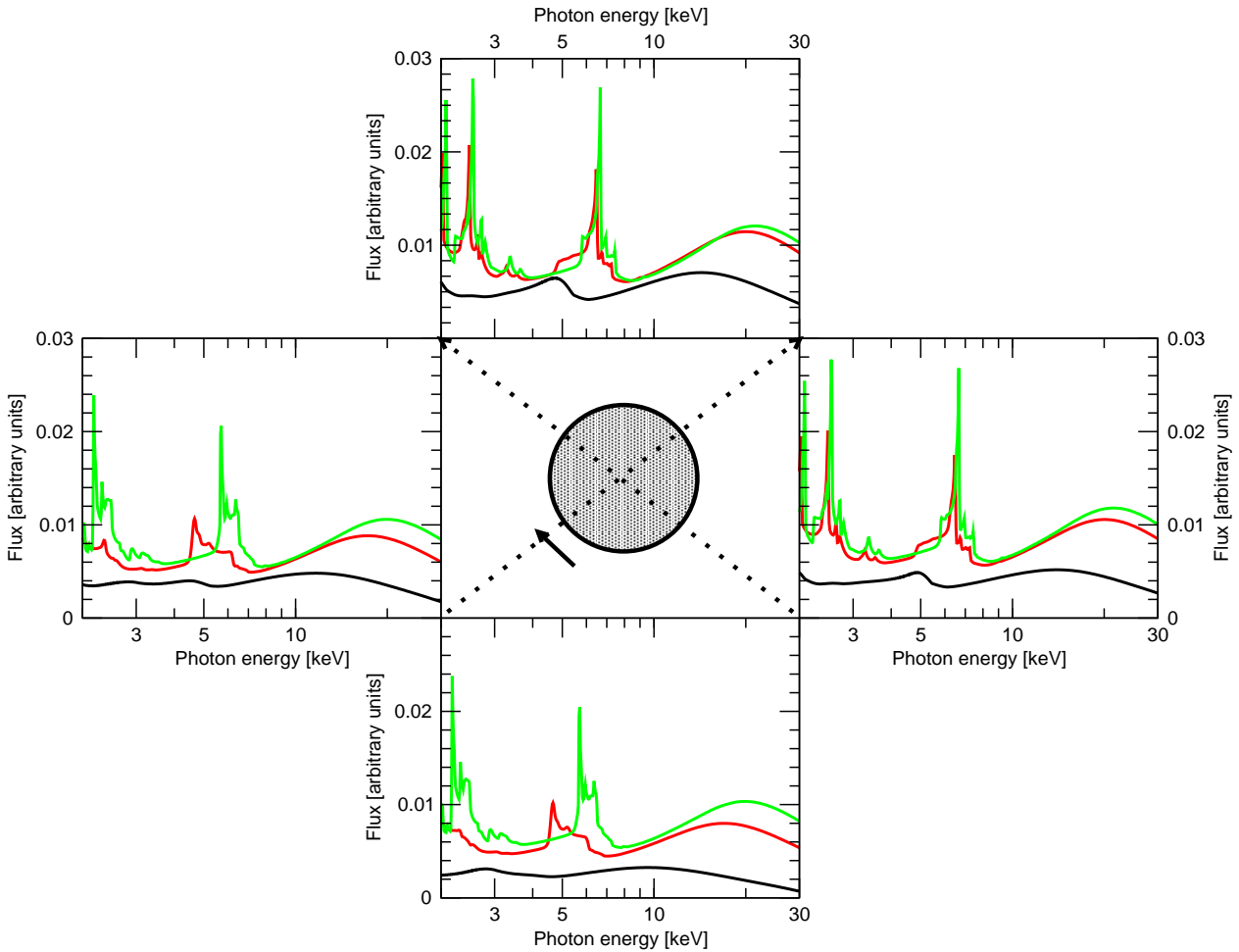}
  \caption{Time-integrated spectra seen by a distant observer for a hot spot
  completing a quarter of an orbit. The observer is located at the bottom of
  the figure and inclined by $i = 30^\circ$ with respect to the disk
  normal. The bottom figure hence represents the orbital phase $\phi =
  0^\circ$, when the spot is at the closest approach to the
  observer. The disk and the black hole rotate clockwise. The curves
  denote spots at different disk radii: 3~$R_{\rm g}$ (black),
  10~$R_{\rm g}$ (red), and 40~$R_{\rm g}$ (green).}
  \label{fig:fullspec}
\end{figure}

We show the resulting spectra in Fig.~\ref{fig:fullspec}. The
reprocessing features we are particularly interested in are the
Compton hump and the iron K$\alpha$ line complex. In the local frame
of the disk, the Compton hump is situated around 30~keV and the
laboratory energy of the iron line ranges, depending on the ionization
state, between 6.4 and 6.9~keV. The figure shows how the reprocessing
features are shifted by gravitational energy shift and Doppler
effects. The gravitational shift depends only on $R$ while the Doppler
effect also changes with the azimuthal phase. As a result of these two
effects, the iron K$\alpha$ line is smeared out and the smearing
becomes stronger with decreasing $R$. Very close to the black hole, at
$R = 3 \, R_{\rm g}$, the line is not recognizable any
more. Furthermore, the relative flux between the Compton hump and the
iron line maximum changes systematically and decreases with $R$. The
closer to the black hole the spectra originate, the flatter they
appear. Furthermore, light-bending effects and the frame dragging
around the fast-spinning black hole are important. They induce an
asymmetric shape of the relativistic transfer functions \citep[as seen on the
maps by][]{dovciak2004b}. The normalization of the spectrum is
higher between $\phi = 135^\circ$ and $\phi = 315^\circ$ than for the
other half-orbit. Also the deformation of the iron line differs
between the two half-orbits.

\section{Short-term flares for different system parameters}

Next, we consider short-term flares. Their irradiation time is
supposed to be comparable to the light-crossing time of the hot
spot. Therefore, the spot re-emission evolves from the spot center, where the
first incident photons arrive, to the border. The local lightcurves show a
rising phase, a maximum,  and a fade-out phase. We compute models
changing the inclination of the observer and the global parameters of
the accreting black hole. We set $a/M = 0$ and calculate reprocessing
spectra for spots at disk radii of $R =  7 \, R_{\rm g}$ and $R = 18
\, R_{\rm g}$.

In Table~\ref{tab:specdata} we show hard/soft-flux ratios $\zeta_{\rm
hs} = F$[28 keV]/$F$[4 keV] and equivalent widths of the iron K$\alpha$
line for both the locally emitted and the far-away observed spectra of
a spot passing ``behind'' the black hole ($\phi = 180^\circ$). The
spectra are taken at the maximum of the lightcurve. For almost all
local spectra we examine, the spectral hardness decreases with
increasing inclination. After including the relativistic and Doppler
effects, this trend is reversed for the closer-in hot-spots at $7 \,
R_g$. Generally, the equivalent width seen by a distant observer,
$EW_{\rm far}$(K$\alpha$), is diminished in comparison to the local
value, $EW_{\rm loc}$(K$\alpha$), although the definition of the
continuum level was made by eye, which induces uncertainties. The set
of parameters for $M = 10^7 \, M_\odot$ corresponds to the
characteristics of the Seyfert galaxy MCG-6-30-15 during a strong
flare \citep{ponti2004}. In this case the ratio $F_{\rm X}/F_{\rm
disk}$ is weaker than before, which slightly hardens the reprocessed
spectrum. Adding dilution by the primary radiation (primary~=~1 in
Table~\ref{tab:specdata}, primary~=~0 means pure reflection)
decreases the equivalent widths of the K$\alpha$-line, as expected.

\begin{table}[h]
  \centering
  \caption{Hard/soft X-ray slopes and equivalent widths of the iron
   K$\alpha$-line for X-ray flares passing behind a black hole ($\phi =
   180^\circ$). The spectra at the peak of the lightcurves are
   considered. Units: $M$ in $M_\odot$, $\dot m_{\rm disk}$ in
   Eddington units, $R$ in $R_{\rm g}$, and equivalent widths in eV.}
  \label{tab:specdata}
  \vskip 0.5 truecm
  {\small
  \begin{tabular}{cccccccccc}
    \hline
    \noalign{\smallskip}
    $M$ & $\dot m_{\rm disk}$ & $R$ & $F_{\rm X}/F_{\rm disk}$ & primary & $i$
    & $\zeta_{\rm hs}^{\rm loc}$ & $EW_{\rm loc}$(K$\alpha$) & $\zeta_{\rm
    hs}^{\rm far}$ & $EW_{\rm far}$(K$\alpha$)\\ 
    \noalign{\smallskip}
    \hline
    \noalign{\smallskip}
      $10^8$ & 0.001 & 18 & 144 & 0 & $10^\circ$ & 0.46 & 1380 & 0.94 & 1595\\
      $10^8$ & 0.001 & 18 & 144 & 0 & $30^\circ$ & 0.46 & 1570 & 0.94 & 1605\\
      $10^8$ & 0.001 & 18 & 144 & 1 & $30^\circ$ & 0.20 &  250 & 0.21 &   80\\
      $10^8$ & 0.001 & 18 & 144 & 0 & $60^\circ$ & 0.39 & 1580 & 0.88 & 1595\\
      $10^8$ & 0.001 &  7 & 144 & 0 & $10^\circ$ & 0.72 & 2030 & 1.10 & 1730\\
      $10^8$ & 0.001 &  7 & 144 & 0 & $30^\circ$ & 0.73 & 2070 & 1.12 & 1745\\
      $10^8$ & 0.001 &  7 & 144 & 1 & $30^\circ$ & 0.21 &  245 & 0.29 &  135\\
      $10^8$ & 0.001 &  7 & 144 & 0 & $60^\circ$ & 0.63 & 1950 & 1.16 & 1700\\
      $10^7$ &  0.02 & 18 &  9  & 0 & $10^\circ$ & 0.59 & 1890 & 1.87 & 1865\\
      $10^7$ &  0.02 & 18 &  9  & 0 & $30^\circ$ & 0.58 & 1880 & 1.83 & 1845\\
      $10^7$ &  0.02 & 18 &  9  & 1 & $30^\circ$ & 0.21 &  255 & 0.21 &   70\\
      $10^7$ &  0.02 & 18 &  9  & 0 & $60^\circ$ & 0.51 & 1740 & 1.67 & 1765\\
    \noalign{\smallskip}
    \hline
  \end{tabular}
 }
\end{table}

\section{Discussion}

There are good arguments to assume that the X-ray irradiation pattern
across an AGN accretion disk is not azimuthally symmetric. Indications
of this asymmetry can be found for NGC~3516 in \citet{iwasawa2004},
for NGC~5548 in \citet{kaastra2004}, or for Mrk~766 in
\citet{turner2006}. The authors use timing analysis of spectral data
from {\it XMM-Newton} and {\it Chandra} to constrain orbital motions
of localized X-ray sources (hot spots). More constraints on the
azimuthal irradiation pattern of the accretion disk are expected from
broader spectral coverage including the region of the Compton
hump. The analysis presented here shows that some systematic
dependencies of the reprocessed spectrum exist: A promising result is
that the normalization of the observed Compton hump with respect to
the soft X-ray continuum flattens toward the disk center at all
orbital phases. Additionally, the relativistic effects induce a
phase-modulation of the reprocessing features. For a spot orbiting at
$R = 10 \, R_{\rm g}$, the deviation of the Compton hump maximum from
its orbital average is up to 18\%. Thereby, the hump's centroid shifts
by up to 8\%. In principle, observations from {\it Suzaku} and later
from {\it Simbol-X}, {\it XEUS}, or {\it Constellation-X} can
therefore further constrain position and motion of X-ray emitting
spots co-orbiting with the disk.

By quantifying the re-processed spectra for a specific azimuthal phase
of the orbiting hot spot we find that strong dependencies of
$\zeta_{\rm hs}^{\rm far}$ and $EW_{\rm far}$(K$\alpha$) exist also on
the global black hole parameters and on the intensity of the
flare. Therefore it remains a multi-parameter problem to constrain
the black hole properties and the irradiation pattern of the
accretion disk. The consideration of variability data is a
possibility to face this problem. Investigations of the power spectral
density of AGN can successfully constrain the black hole mass
\citep[e.g.][]{mchardy2005}. We also expect further constraints on the
black hole spin and on the irradiation pattern of the disk from our {\it
rms} modeling \citep{goosmann2006a} when applied to a broader spectral
range. 

\vspace{0.3cm}

\noindent {\it Acknowledgements.} We are grateful to Anne-Marie Dumont and
Agata R{\' o}{\. z}a{\' n}ska for their help computing the local spectra and
the vertical disk profiles.


\begin{thebibliography}{}
\setlength{\itemsep}{0cm}

{\footnotesize

\bibitem[Czerny et al.(2004)]{czerny2004} Czerny, B., R{\'o}{\.z}a{\'n}ska,
A., Dov{\v c}iak, M., Karas, V., \& Dumont, A.-M.\ 2004, A\&A, 420, 1 

\bibitem[Dov{\v c}iak(2004)]{dovciak2004b} Dov{\v c}iak, M. \ 2004,
PhD-thesis, Charles University Prague, astro-ph/0411605

\bibitem[Dov{\v c}iak et al.(2004)]{dovciak2004a} Dov{\v c}iak, M.,
Karas, V., \& Yaqoob, T.\ 2004, ApJS, 153, 205

\bibitem[Dumont et al.(2000)]{dumont2000} Dumont, A.-M., Abrassart,
A., \& Collin, S.\ 2000, A\&A, 357, 823

\bibitem[Dumont et al.(2003)]{dumont2003} Dumont, A.-M., Collin, S.,
Paletou, F., Coup{\' e}, S., Godet, O., \& Pelat, D.\ 2003, A\&A, 407,
13

\bibitem[Galeev et al.(1979)]{galeev1979} Galeev, A.~A., Rosner, R.,
\& Vaiana, G.~S.\ 1979, ApJ, 229, 318

\bibitem[Haardt et al.(1994)]{haardt1994} Haardt, F., Maraschi, L., \&
Ghisellini, G.\ 1994, ApJL, 432, L95

\bibitem[Haardt \& Maraschi(1991)]{haardt1991} Haardt, F., \& 
Maraschi, L.\ 1991, ApJL, 380, L51 

\bibitem[Iwasawa et al.(2004)]{iwasawa2004} Iwasawa, K., Miniutti, 
G., \& Fabian, A.~C.\ 2004, MNRAS, 355, 1073

\bibitem[Goosmann(2006)]{goosmann2006b} Goosmann, R.~W. \ 2006, PhD
  thesis, Universit\"at Hamburg

\bibitem[Goosmann et al.(2006)]{goosmann2006a} Goosmann, R.~W., 
Czerny, B., Mouchet, M., Ponti, G., Dov{\v c}iak, M., Karas, V., 
R{\'o}{\.z}a{\'n}ska, A., \& Dumont, A.-M.\ 2006, A\&A, 454, 741 

\bibitem[Kaastra et al.(2004)]{kaastra2004} Kaastra, J.~S., et al.\ 
2004, A\&A, 422, 97 

\bibitem[McHardy et al.(2005)]{mchardy2005} McHardy, I.~M., Gunn, 
K.~F., Uttley, P., \& Goad, M.~R.\ 2005, MNRAS, 359, 1469 

\bibitem[Ponti et al.(2004)]{ponti2004} Ponti, G., Cappi, M., 
Dadina, M., \& Malaguti, G.\ 2004, A\&A, 417, 451 

\bibitem[R{\' o}{\. z}a{\' n}ska et al.(2002)]{rozanska2002} R{\'
o}{\. z}a{\' n}ska, A., Dumont, A.-M., Czerny, B., \& Collin, S.\
2002, MNRAS, 332, 799

\bibitem[Turner et al.(2006)]{turner2006} Turner, T.~J., Miller, 
L., George, I.~M., \& Reeves, J.~N.\ 2006, A\&A, 445, 59 

}

\end{thebibliography}
\end{document}